\documentclass[]{spie}  
\usepackage[dvips]{graphicx}
\usepackage{epsfig}
\usepackage{mystyle}

\title{BRIGHT lights, BIG city:\\
Massive galaxies, giant Ly-a nebulae, and proto--clusters}

\author{Wil van Breugel\supit{a}, Michiel Reuland\supit{a,c}, Wim de Vries\supit{a}, Adam Stanford\supit{a}, Arjun Dey\supit{b} \\
Jaron Kurk\supit{c}, Bram Venemans\supit{c}, Huub R\"ottgering\supit{c}, George Miley\supit{c}, Carlos De Breuck\supit{d} \\
Mike Dopita\supit{e}, Ralph Sutherland\supit{e}, Joss Bland--Hawthorn\supit{f} 
\skiplinehalf 
\supit{a}University of California - Lawrence Livermore National Laboratory \\ 
P.O. Box 808, Mailstop L-413, Livermore CA 94551, U.S.A.
\\
\supit{b} National Optical Astronomy Observatories \\
950 N. Cherry Ave, Tucson, AZ 85719 \\
\supit{c} Leiden Observatory \\
P.O. Box 9513, NielsBohrweg 2, Leiden 2300 RA, The Netherlands \\
\supit{d} Inst. d'Astrophysique de Paris \\
98bis Boulevard Arago, 75014 Paris, France \\
\supit{e} Australian National University Res. School of Astronomy \& Astrophysics \\
Cotter Road, Weston Creek, ACT 2611, Australia \\
\supit{f} Anglo--Australian Observatory \\
P.O. Box 296, Epping, NSW 2121, Australia
}

\authorinfo{Further author information: (Send correspondence to
W.v.B.)\\W.v.B.: E-mail: wil@igpp.llnl.gov}

\begin{document} 
  \maketitle 

\begin{abstract}
High redshift radio galaxies are great cosmological tools for pinpointing
the most massive objects in the early Universe: massive forming galaxies,
active super--massive black holes and proto--clusters. We report on
deep narrow--band imaging and spectroscopic observations of several $z >
2$ radio galaxy fields to investigate the nature of giant Ly--$\alpha$
nebulae centered on the galaxies and to search for over--dense regions
around them. We discuss the possible implications for our understanding
of the formation and evolution of massive galaxies and galaxy clusters.

\end{abstract}


\keywords{High redshift, radio galaxies, massive galaxies, proto--clusters,
black holes}



\def\cf{{c.f.,~}}
\def\ie{{i.e.,~}}
\def\eg{{e.g.,~}}
\def\etal{{et al.~}}
\def\tn1338{{TN~J1338$-$1942}}
\def\mrc0943{{MRC~0943$-$242}}
\def\wn0305{{WN~J0305+3525}}
\def\hzrgs{HzRGs}
\def\hzrg{HzRG}
\def\alp{\alpha^{20}_{90}}

\def\lya{\ifmmode {\rm Ly\alpha}\else{\rm Ly$\alpha$}\fi}
\def\Ha{\ifmmode {\rm H\alpha}\else{\rm H$\alpha$}\fi}
\def\Lya{Ly$\alpha$}
\def\CIV{\hbox{C~$\rm IV$}~$\lambda$~1549}
\def\HeII{\hbox{He~$\rm II$}~$\lambda$~1640}
\def\OII{[\hbox{O~$\rm II$}]~$\lambda$~3727}
\def\OIII{[\hbox{O~$\rm III$}]~$\lambda$~5007}
\def\deg{$^{\circ}$}
\def\arcmin{\ifmmode \rlap.{^{\prime}}\else
    $\rlap.{^{\prime}}$\fi}
\def\spose#1{\hbox to 0pt{#1\hss}}
\def\simlt{\mathrel{\spose{\lower 3pt\hbox{$\mathchar"218$}}
     \raise 2.0pt\hbox{$\mathchar"13C$}}}
\def\simgt{\mathrel{\spose{\lower 3pt\hbox{$\mathchar"218$}}
     \raise 2.0pt\hbox{$\mathchar"13E$}}}

\def\OmM{\ifmmode {\Omega_{\rm M}}\else $\Omega_{\rm M}$\fi}
\def\OmL{\ifmmode {\Omega_{\Lambda}}\else $\Omega_{\Lambda}$\fi}
\def\ph{\ifmmode {h_{65}^{-1}}\else $h_{65}^{-1}$\fi}
\def\psqh{\ifmmode {h_{65}^{-2}}\else $h_{65}^{-2}$\fi}
\def\kpc{{\rm\,kpc}}
\def\kms{\ifmmode {\rm\,km\,s^{-1}}\else ${\rm\,km\,s^{-1}}$\fi}
\def\kmps{\ifmmode {\rm\,km~s^{-1}} \else ${\rm\,km\,s^{-1}}$\fi}
\def\dynpcm{\ifmmode {\rm\,dyn\,cm^{-2}} \else {${\rm\,dyn\,cm^{-2}}$}\fi}
\def\ergps{\ifmmode {\rm\,erg\,s^{-1}} \else {${\rm\,erg\,s^{-1}}$}\fi}
\def\ergpspcm2{\ifmmode {\rm\,erg\,s^{-1}\,cm^{-2}} \else 
{${\rm\,erg\,s^{-1}\,cm^{-2}}$}\fi}

\def\msun{\ifmmode {\rm\,M_\odot}\else ${\rm\,M_\odot}$\fi}
\def\Msun{\ifmmode {\rm\,M_\odot}\else ${\rm\,M_\odot}$\fi}  
\def\lsun{\ifmmode {\rm\,L_\odot}\else ${\rm\,L_\odot}$\fi}
\def\Lsun{\ifmmode {\rm\,L_\odot}\else ${\rm\,L_\odot}$\fi}
\def\rsun{\ifmmode {\rm\,R_\odot}\else ${\rm\,R_\odot}$\fi}
\def\Rsun{\ifmmode {\rm\,R_\odot}\else ${\rm\,R_\odot}$\fi}
\def\Msunpyr{\ifmmode {\rm\,M_\odot\,yr^{-1}} \else
{${\rm\,M_\odot\,yr^{-1}}$}\fi}

\section{Radio sources as cosmological probes}
\label{sect:why}  

High redshift radio galaxies (\hzrgs; $z > 2$) are great beacons for
pinpointing the most massive objects in the early universe, whether these
are galaxies, black holes or even clusters of galaxies.  At {\it low}
redshifts powerful, non--thermal radio sources are uniqely associated
with massive ellipticals. Their twin--jet, double--lobe morphologies
and large luminosities suggested already early on that such galaxies
must also have spinning, super--massive black holes (SMBH's) in
their centers\cite{Rees78,Blandford82}.  We now know that the masses
of the stellar bulges of galaxies and their central black holes are
correlated\cite{Magorrian98,Gebhardt00,Ferrarese00}, suggesting a causal
connection. If radio sources are powered by SMBH's then it is no longer
a surprise that their parent galaxies occupy the upper end of the galaxy
mass function.

There is excellent evidence that radio galaxies are also the most massive
systems at {\it high} redshifts, even though their parent galaxies
are very young and may still be forming.  The combined near--infrared
`Hubble' $K-z$ relation for radio and field galaxies\cite{DeBreuck02a}
shows that \hzrgs\ are among the most luminous systems at any given
epoch up to $z\sim 5$. Between $0 < z < 2.5$ this $K - z$ diagram
can be modeled using passive evolution of 5$L_\star$
ellipticals\cite{Jarvis01} which formed at $5 < z < 10$.  Other evidence
includes the direct detection of absorption lines from massive
young stars\cite{Dey97}, large ($\sim 50 - 70$ kpc), multi--component
rest--frame UV and optical morphologies\cite{Pentericci99}, the 'hyper'
luminous rest--frame far--infrared luminosities ($L_{FIR} \simgt 10^{13}
\Lsun$) and huge implied star formation rates\cite{Archibald01,Reuland02}
($\simgt 2000 \Msunpyr$), 
and last but not least, the very extended
($\sim 30 - 50$ kpc) molecular gas and dust clouds that have recently
been discovered\cite{Papadopoulos00,DeBreuck02b}  
around several \hzrgs\ 
showing that the star formation occurs on galaxy wide scales.

\hzrgs\ are also excellent tools for finding over--dense regions 
(`proto--clusters') at high redshift. This is because, in standard Cold Dark
Matter (CDM) scenarios, galaxy formation is a highly `biased' process: the
most massive galaxies, and the largest clusters of galaxies, are expected
to emerge from regions with the largest over--densities\cite{Kaiser84}.
Simply put: the most massive systems (galaxies, SMBH's and galaxy clusters)
hang out together, and radio sources are a great way to find
them and to investigate their interrelations and evolution.

In this paper we report on deep narrow--band imaging and spectroscopic
observations of several $z > 2$ radio galaxy fields using the Keck
and ESO/VLT telescopes to investigate the nature of giant Ly--$\alpha$
nebulae centered on the galaxies and to search for over--dense regions
around them. We discuss the possible implications for our understanding
of the formation and evolution of massive galaxies and galaxy clusters.
We will adopt the cosmological parameters $\OmM = 0.3$, $\OmL = 0.7$,
$H_{0} = 65 \kmps\,\rm Mpc^{-1}$, for which the age of the Universe
at $z \sim$ 2, 3 and 4 is 3.5, 2.3 and 1.6 Gyr respectively, and the
angular--to--linear transformations are 9.0, 8.3 and 7.5 \kpc\ arcsec$^{-1}$.

\section{Giant \lya\ Nebulae} \label{sect:lya}

Since the discovery of the first \hzrgs\ it has been
known that they are surrounded by large \lya\ emission--line
nebulae\cite{Eisenhardt92,Chambers90,vanOjik96,Kurk02}.  Because the
nebulae are centered on large, young galaxies they provide a unique
opportunity to study how such galaxies form (accretion or merging ?),
and about AGN/starburst feedback and chemical enrichment during this
process (heating or outflow ?). We therefore observed the \lya\ nebulae of
several \hzrgs\ and describe the results for two of them in detail below.
A summary of the largest high redshift \lya\ nebulae currently known is
given in Table 1.

\begin{table}[ht]
\centering
\caption{Large, High Redshift \lya\ Nebulae}
\begin{tabular}{llllll}
\hline
\hline
Name & z & Size                    & Log(Ly-$\alpha$)$^b$ & Telescope & Reference \\
     &   & $\kpc \times \kpc$      & \ergps           &           &           \\
\hline
\hline
MRC~1138$-$262   & 2.156 & 150 $\times$ 120 & 44.3 & VLT 8--m     & Ref.~\citenum{Kurk02} \\
Francis' Blob$^a$ & 2.380 & 100 $\times$  30 & 43.9 & CTIO 4--m   & Ref.~\citenum{Francis01} \\
Steidel Blobs$^a$ & 3.09 & 140 $\times$ 120 & 44.1 & Palomar 5--m & Ref.~\citenum{Steidel00} \\
B2~0902+34       & 3.395 & 140 $\times$ 100 & 44.8 & Keck 10--m   & Ref.~\citenum{Reuland02} \\
4C~1243+036$^c$  & 3.570 & 160 $\times$ 50  & 44.0 & ESO 4--m     & Ref.~\citenum{vanOjik96} \\
4C60.07          & 3.791 & 110 $\times$ 60  & 45.1 & Keck 10--m   & Ref.~\citenum{Reuland02} \\
4C41.17          & 3.798 & 190 $\times$ 130 & 45.2 & Keck 10--m   & Ref.~\citenum{Reuland02} \\
TN~J1338$-$1942  & 4.102 & 150 $\times$ 40  & 44.7 & VLT 8--m     & Ref.~\citenum{Venemans02} \\
\hline
\end{tabular}
\begin{tabular}{p{11cm}}
\quad $^a$ Radio 'quiet' \\
\quad $^b$ Depends on surface brightness limit
(larger telescopes detect larger Ly--$\alpha$ nebulae) \\
\quad $^c$ Maximum size determined from long slit spectrocopy, not imaging
\end{tabular}
\end{table}

\subsection{4C41.17 at $z = 3.798$}

\begin{figure}[t]
\begin{center}
\plottwo{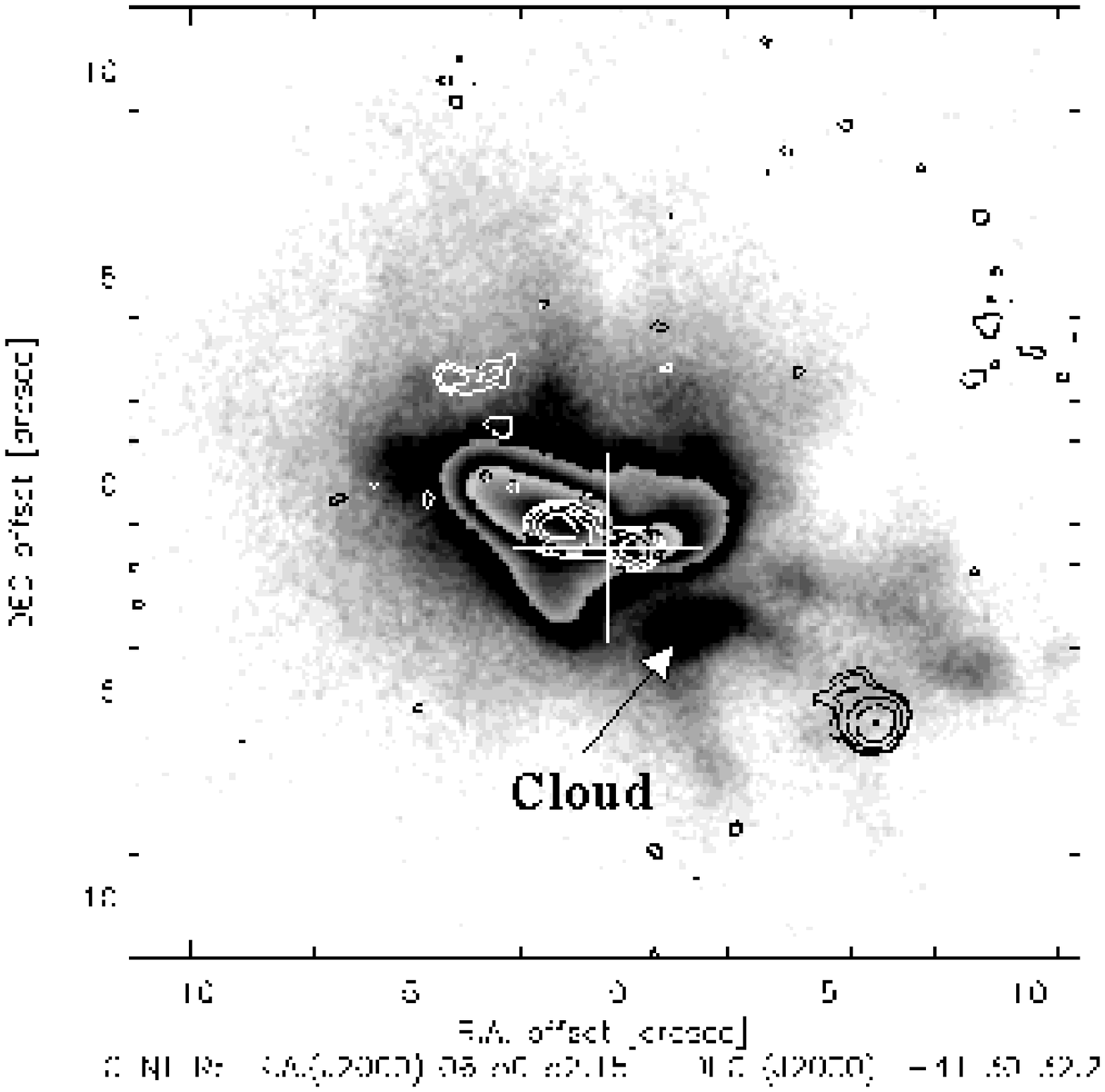}{0.45}{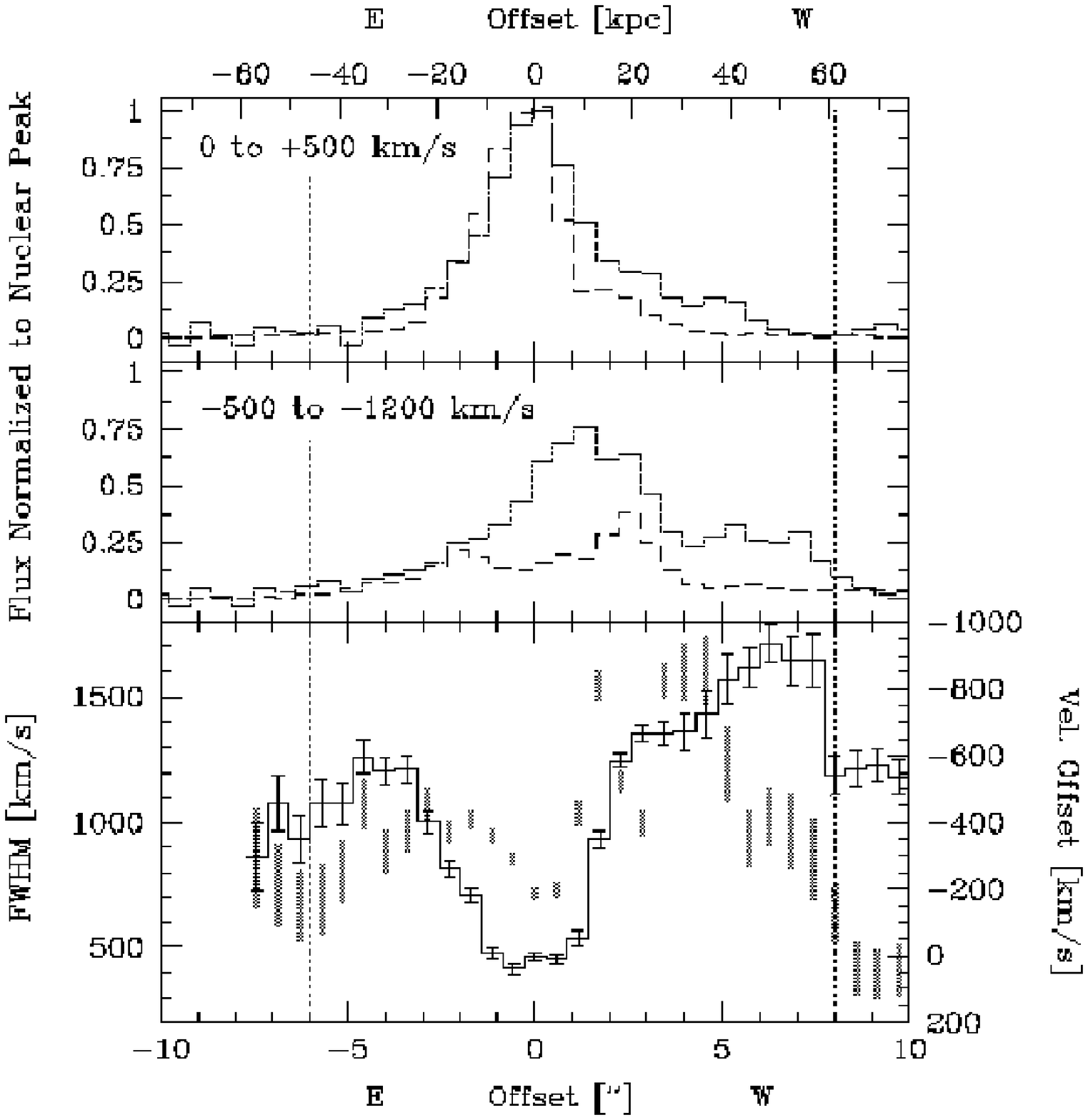}{0.45}
\caption{
[Left] Keck narrow--band \lya\ image (grey scale) 
of 4C41.17 overlaid with a 6~cm VLA radio image (contours)\cite{Reuland02}.
The cross shows the
position of the radio core.
[Right] Normalized surface brightness profiles of the [OII] (solid)
and \lya\ (dashed) emission along the inner radio axis and the long
south--west filament. Top: Extended [OII] in the
velocity range $0$ to $+500$\kms.  The spatial zero--point corresponds
to the position of the radio core.  Middle: Similar to top panel but for
the velocity range $-500$ to $-1200$\kms\ (Note: The $0$ to $-500$\kms\
range for [OII] is affected by near--infrared sky lines and is not shown).
Bottom: Relative velocity (solid line) and velocity dispersion (bars)
of the \lya\ emission. Bar and symbol size indicate the respective
uncertainties in the individual measurements. The projected distances
of the south--west and north--east radio lobes along the slit direction
are indicated (dotted lines).
\label{fig:fig1}}
\end{center}
\end{figure}

4C41.17 was imaged using a custom--made, high throughput interference
filter with a 65~\AA\ bandpass centered at the redshifted \lya\ =
5839~\AA\ line with the Echellette Spectrograph and Imager\cite{Sheinis00}
at the Cassegrain focus of the Keck II 10m telescope.  The data
were obtained during photometric conditions and good seeing (FWHM =
0.57\arcsec). With a total exposure time of 27,600\,s this \lya\ image is
the most sensitive obtained to date and reaches a surface brightness of
$8.0 \times 10^{-19} \rm \,erg\,s^{-1}\,cm^{-2}\,arcsec^{-2}$ (3$\sigma$
limit in a 2.0\arcsec\ diameter aperture).  A broad--band $R$ image,
obtained as part of a multi--band spectral energy distribution study
of the 4C41.17 field was scaled and subtracted from the \lya\ image to
construct the pure emission--line image shown in Figure 1.  With linear
dimensions of $190 \kpc \times 130 \kpc$ at the detection limit the
4C41.17 nebula is the largest presently known (Table 1).

A striking morphological feature of the nebula is a cone--shaped
structure emanating from the center of the galaxy, with a 75\kpc\ long
radial filament in the vicinity of the south--west radio lobe and a
crescent--shaped cloud with radial horns.  This morphology resembles
that of other, nearby active and starburst galaxies and is suggestive
of entrainment by the radio source\cite{vanBreugel86}, outflow driven
by radiative pressure from the AGN\cite{Dopita02}, or a starburst
superwind\cite{Lehnert99}.

Since \lya\ is a resonance line, it is important to determine whether the
nebula is ionized, or rather, results from the scattering of the \lya\
photons (produced near the nucleus) by an extended neutral hydrogen
gaseous halo.  We therefore obtained a long--slit spectrum of 4C41.17
with the Keck II near--infrared spectrograph\cite{McLean98} (NIRSPEC) to
measure the extent and kinematics of other lines, \OII\ and \OIII, along
the brighest filament of the \lya\ nebula.  We discovered both extended
[OII] and [OIII] emission. In particular, [OII] emission was detected
as far as $\sim 60$\kpc\ from the nucleus (Fig.  1).  This shows that
the \lya\ nebula, at least in this direction, is not due to scattering
but must be locally ionized. Furthermore, the presence of the emission
lines of oxygen shows that the halo gas is not chemically pristine
(primordial H, He) gas but has been enriched either at much earlier
epochs (during the `Dark Ages'), or by more recent star formation.

The near--infrared [OII] and optical \lya\ kinematics show that both
emission lines exhibit large blue--shifted velocities $ \sim 600 -
900\kms\ $ (in projection) along the radio axis.  In particular,
the gas is very disturbed along the south--west filament, with \lya\
velocity widths ranging up to $\Delta v_{F} \sim 900 - 1600 \kms\ $.
Beyond the radio hotspot, the velocity and velocity widths decrease
abruptly. Evidence for entrainment of emission--line gas in
long filaments, even if at considerable distance from the observed
radio emission, has also been seen in nearby radio galaxies \ie
4C29.30\cite{vanBreugel86}.

We note that in the canonical picture of radio sources the hotspots and
lobes are surrounded by bowshocks and cocoons of radio quiet, shock
heated gas with a scale sizes wich are significantly larger than the
observed radio emission\cite{Carilli88,Begelman89}. In this picture the
emission--line filaments located at the interface of the cocoons and
the ambient gas and are not in direct contact with the radio lobes or
hotspots and can even extend {\it beyond} radio 
hotspots\cite{Maxfield02} (Figs 1, 2), 
in particular if one also accounts for projection effects
and the fact that the radio observations only show the highest surface
brightness regions.  The kinematics along the filament in 4C41.17, its
radial filamentary structure, and the chemical enrichment of this gas
therefore all indicate a process of entrainment of material away from the
central regions by the radio source. The outer regions of \lya\ nebulae
may not have been affected by the sources, especially if they are still
young and small, and one would expect these regions to be less enriched,
as has been observed\cite{Binette00,Jarvis02}.

\subsection{4C60.07 at $z = 3.790$}

\begin{figure}[t]
\begin{center}
\plottwo{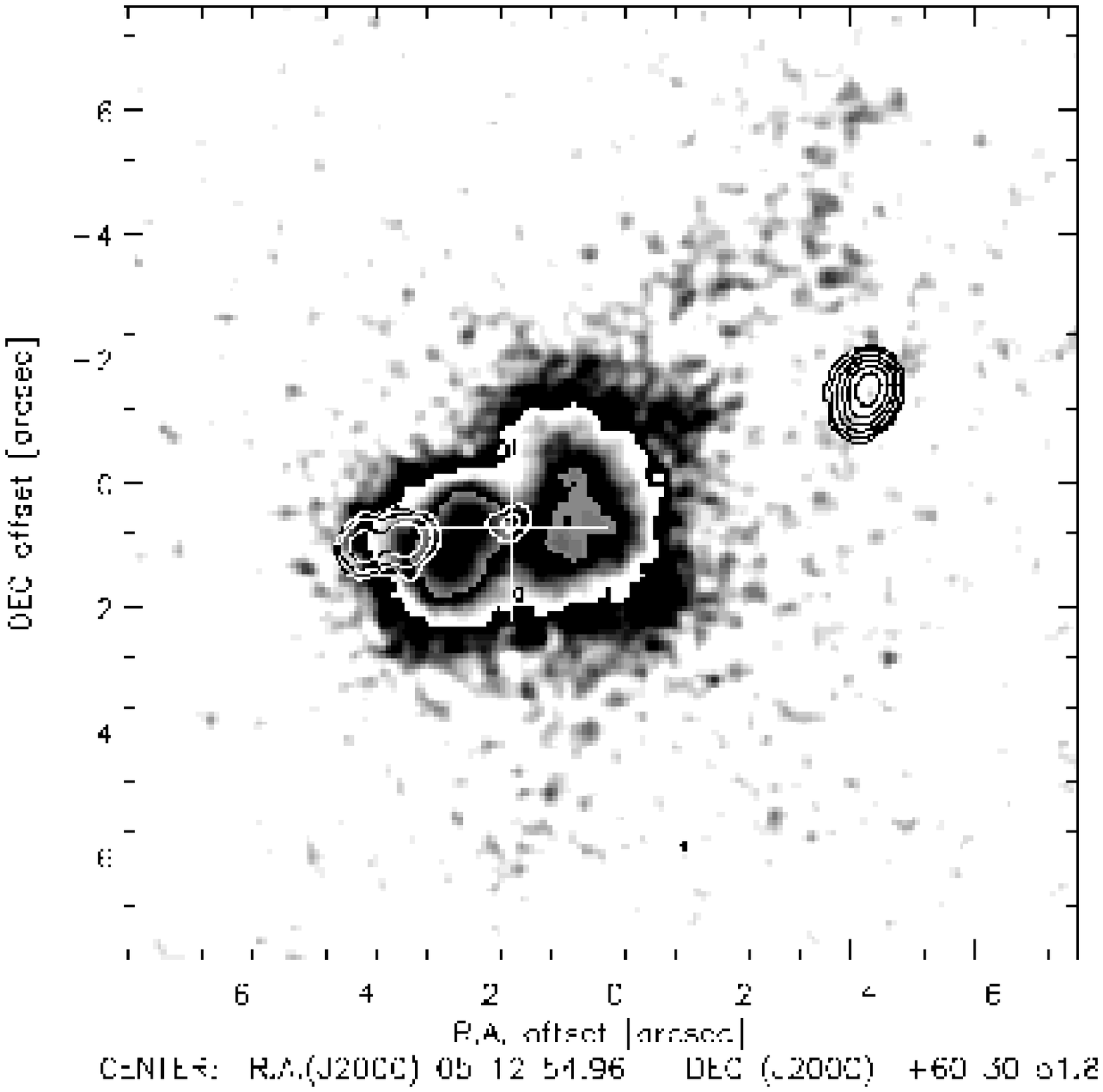}{0.45}{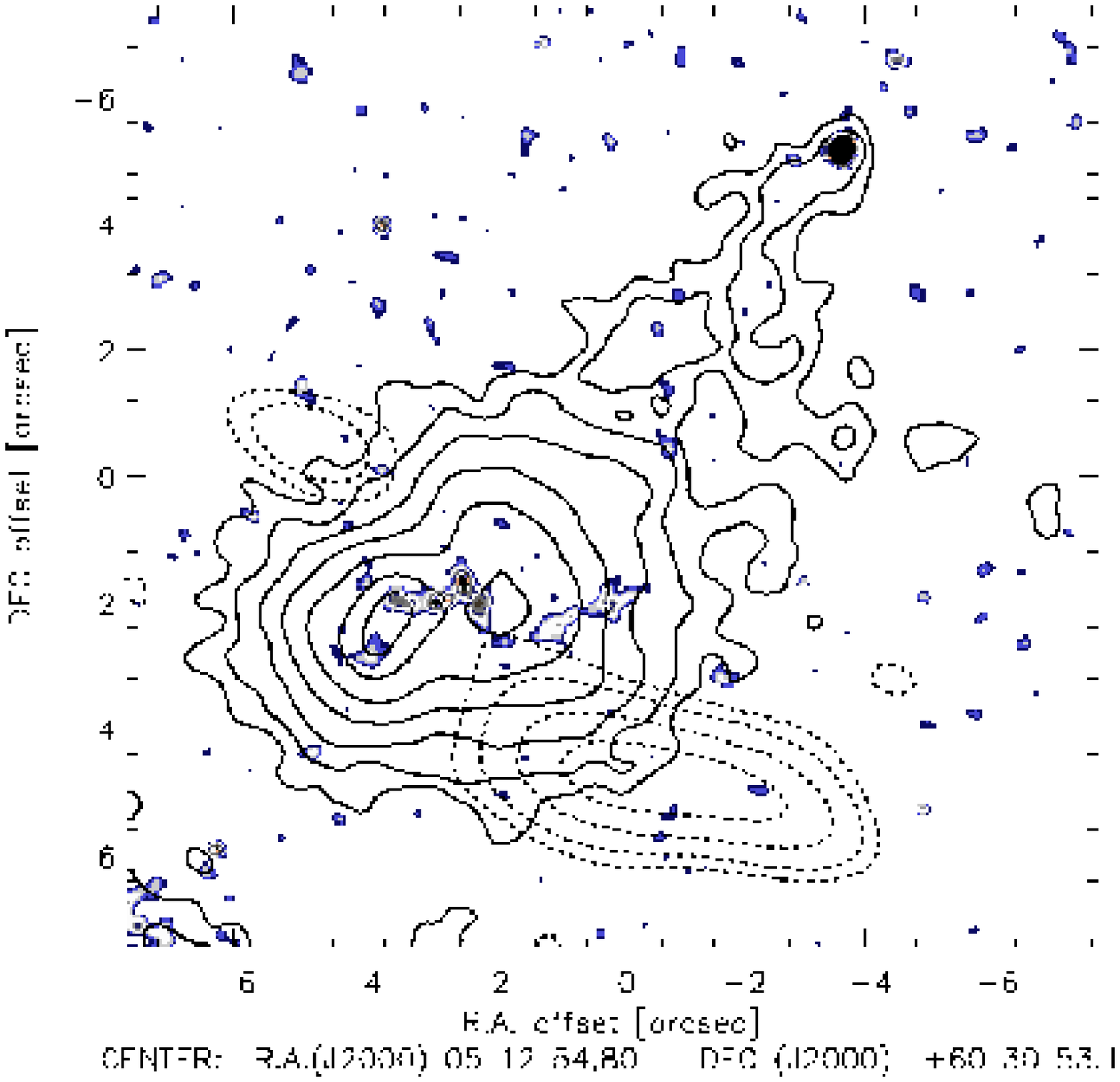}{0.45}
\caption{
[Left] Keck narrow--band \lya\ image\cite{Reuland02} (grey scale) of
4C60.07 overlaid with a VLA radio image (contours) at 6~cm
(contours).  The cross shows the position of the radio core.
[Right] HST `R'--band image of 4C60.07 (grey scale) overlaid with a Keck
narrow--band \lya\ image (contours) and IRAM interfermometer dust image\cite{Reuland02}
(dotted lines).
\label{fig:fig2}}
\end{center}
\end{figure}

Because of its very similar redshift the radio galaxy 4C60.07 could be
observed with the same narrow--band filter as 4C41.17. The total observing
time on source was 7,200\,s, resulting in a surface brightness detection
limit of $\sim 2 \times 10^{-18} \rm \,erg\,s^{-1}\,cm^{-2}\,arcsec^{-2}$
(3$\sigma$ limit in a 2.0\arcsec\ diameter aperture).  Figure 2 shows
the \lya\ nebula in relationship to the radio source, rest--frame UV
HST structure, and the extended dust and molecular gas that has been
found in this system\cite{Reuland02,Papadopoulos00}.
Unfortunately the \lya\ image is not as deep as for 4C41.17, not only
because of the shorter intergration time, but also because 4C60.07 has
significant Galactic foreground extinction ($\sim 1.6$ magnitude at the
central wavelength of the narrow band filter).

Nevertheless one notes several similarities to 4C41.17 and other radio
galaxy emission--line nebulae.  First, the brightest \lya\ emission
is associated with the radio lobe which is closest to the AGN. This
is presumably because of asymmetries in gas density distribution, with
denser gas providing more rampressure on the hotspots of the expanding
radio source\cite{McCarthy87}.  Second, the galaxy structure is
very clumpy and approximately aligned with the radio source. This may
be due to induced star formation by radio lobes which expand sideways
into the cold, dense gas seen in the mm--maps. Good evidence for 
this has been found in 4C41.17\cite{Dey97,Bicknell00}. 
Third, and this is the first such evidence, the \lya\
and dust/molecular gas emission show little overlap. We don't know how
common this is since to date only three \hzrgs\ have been imaged with mm
interferometers with sufficient sensitivity to map their cold gas and dust
emission and of these only 4C60.07 has a deep \lya\ image. However, an
anti--correlation between \lya\ and cold gas certainly is not suprising 
given that \lya\ is easily scattered and absorbed. In fact the overall
impression is that the \lya\ emission is more or less orthogonal to the
dust and cold gas distribution, resembling M82 and its emission--line
starburst superwind\cite{Lehnert99}.

\subsection{Conclusions for Ly--$\alpha$ nebulae}

Since many \hzrgs\ seem to be surrounded by giant \lya\ nebulae it
seems likely that they play an important role in the formation of
massive galaxies.  What is the origin of this gas, what its fate?

With respect to the origin we should note that large \lya\ nebulae also
exist {\it without} central radio galaxies or AGN, and that these also
show large velocity widths and gradients\cite{Steidel00,Francis01}. In the
absence of an obvious central source of ionization and/or outflow (but
see Ref.~\citenum{Chapman01}) this suggests that \lya\ nebulae are due to the accretion
of primordial (?), cooling gas in large CDM halos\cite{Fardal01,Haiman00}.
This gas provides a large reservoir of
galaxies which supplies the building materials for galaxies forming at
their centers.

As to the fate: Starburst\cite{Silk87} and AGN\cite{Silk98} outflows may eject
accreting gas in forming galaxies when such outflows can overcome
the galaxian potential, thereby limiting star formation and galaxy
growth. This feedback mechanism may explain the fixed black hole to bulge
mass ratios observed in nearby galaxies.  In the case of \hzrgs\ our data
shows that this outflow may be significantly helped by shock heating and
entrainment from the radio sources.  We note that radio source induced
'hot bubbles' have also been invoked to reduce gas accretion in the
centers of nearby cooling flow clusters, which seems required to explain
X--ray observations\cite{Nulsen02,Kaiser01}.

Radio sources may be a significant source of heating and chemical
enrichment during their life time\cite{Odea02} ($\sim 10 ^6 - 10^7$ yr). The
cross sections of radio lobes are large\cite{Carilli88}: tens of kpc$^2$
and much larger than the hotspots, which are only the highest surface
brightness features that are visible in radio maps because of the
steep radio lobe spectra and high restframe frequencies.  The multiple
component, asymmetric, and twisted radio structure of 4C41.17 and many
other \hzrgs\ suggests that the central SMBH may experience multiple
periods of radio source activity and precession, presumably triggered by
the interaction with one of the many clumpy components which make up the
complex galaxy structure. If individual clumps move at radial velocities
comparable to the velocity dispersion in the proto--clusters ($\sim 500$
\kms, Table 2) and the closest are within $\sim 10$ \kpc\ of the 
SMBH\cite{Reuland02,Pentericci99} then it takes only $\sim 2 \times
10^7$ yrs to re--supply the AGN. The duty cycles for radio source
activity could therefore be very short, at least during the period that
the galaxies are being assembled. During this period a considerable
fraction of the nebular gas would be heated and expelled.  Not only could
this then be the `end of the beginning' for the galaxy, as its collapse
gives way to mass ejection, it would also be the `end of the beginning'
for the proto--cluster as more and more metals are dispersed throughout
the gas, stimulating cooling, star formation and galaxy evolution.

\section{Proto--clusters}
\label{sect:proto}

\begin{table}[ht]
\centering
\caption{Proto--Clusters}
\begin{tabular}{llllllll}
\hline
\hline
Name & z & N$_{NB}^a$ & N$_{spec}^b$ & $\sigma$ (\kms) & FOV$^c$ & Telescope & Reference \\
\hline
\hline
MRC~1138$-$262 & 2.156 & 51 & 15 & 280, 520 & 35   & VLT 8m     & Ref.~\citenum{Pentericci00} \\
53W002         & 2.39  & 14 &  5 & 530      & 48   & KPNO 4m    & Ref.~\citenum{Keel99} \\
MRC~0943$-$242 & 2.919 & 55 & 18 & 824      & 40   & VLT + Keck & Ref.~\citenum{Venemans03} \\
SSA22$a$       & 3.09  & 72 & 10 & 400      & 77   & Palomar 5m & Ref.~\citenum{Steidel00} \\
4C41.17        & 3.800 & 18 & 3  & --       & 3.5  & Keck 10m   & Ref.~\citenum{Reuland02} \\
TN~J1338$-$1942& 4.102 & 28 & 20 & 325      & 40   & VLT 8m     & Ref.~\citenum{Venemans02} \\
\hline
\end{tabular}
\begin{tabular}{p{11cm}}
\quad $^a$ N$_{NB}$ = Number of Ly--$\alpha$ excess candidates \\
\quad $^b$ N$_{spec}$ =  Number of spectroscopically confirmed Ly--$\alpha$ galaxies \\
\quad $^c$ FOV = field of view in sq.arcmin.
\end{tabular}
\end{table}

Clusters of galaxies are convenient targets for studies of the formation
and evolution of galaxies and may help constrain important cosmological
parameters.  Optical and near--IR surveys, assisted by X--ray selection,
have found clusters\cite{Stanford97,vanDokkum01} 
out to $z \sim 1.3$.  Very
deep optical `Lyman break' surveys have discovered galaxy over--densities
(`proto--clusters') at $z \sim 3.09$\cite{Steidel98,Steidel00}.  Such
observations are time consuming and increasingly difficult at higher
redshifts.  They will not be able to provide substantial numbers of
over--dense regions spread over a large range in redshift, which would
be needed if one wishes to investigate how proto--clusters evolve, at
what redshifts galaxy clusters become virialized, and what the cluster
mass functions are.

A practical way to identify proto--clusters is to
search for over--dense regions around targeted objects.  Radio sources are
ideal for this since they are associated with the most massive galaxies
and, in hierarchical CDM scenarios, therefore live in the most over--dense
regions. Furthermore, radio selection is not biased towards dusty systems
and provides an important alternative method to the optical/near--IR
(rest--frame UV/optical) color selected searches.

The quickest way to identify potential over--dense regions around
\hzrgs\ is to use narrow--band filters, or better even, `tunable'
filters or other such imaging devices (Section 4) if they were available
on the big telescopes. At least some fraction of the young galaxies
in proto--clusters can be expected to be active starforming systems
and such \lya\ excess galaxies (LaEG's) should be detectable in deep
emission--line searches (\lya\ at optical wavelengths for $z > 2$
objects). A deep narrow--band \lya\ image of the $z = 3.09$ SSA22$a$
proto--cluster field, which was initially found using 
broad--band color selection, confirmed this\cite{Steidel00}.

Our Leiden, LLNL and ANU/AAO groups have undertaken a joint program of
deep narrow--band and tunable filter imaging of \hzrg\ fields
with redshifts ranging between $2.2 < z < 5.2$ to identify proto--galaxy
candidates through excess \lya\ emission. This is then followed by
multi--slit spectroscopy to confirm that their redshifts are close to
that of the target \hzrg. This method has been highly successful and we
discuss two recent results in more detail below. A summary of presently
confirmed proto--clusters is shown in Table 2.

\subsection{MRC~0943$-$242 at $z = 2.919$}

\begin{figure}[t]
\begin{center}
\includegraphics[height=10cm]{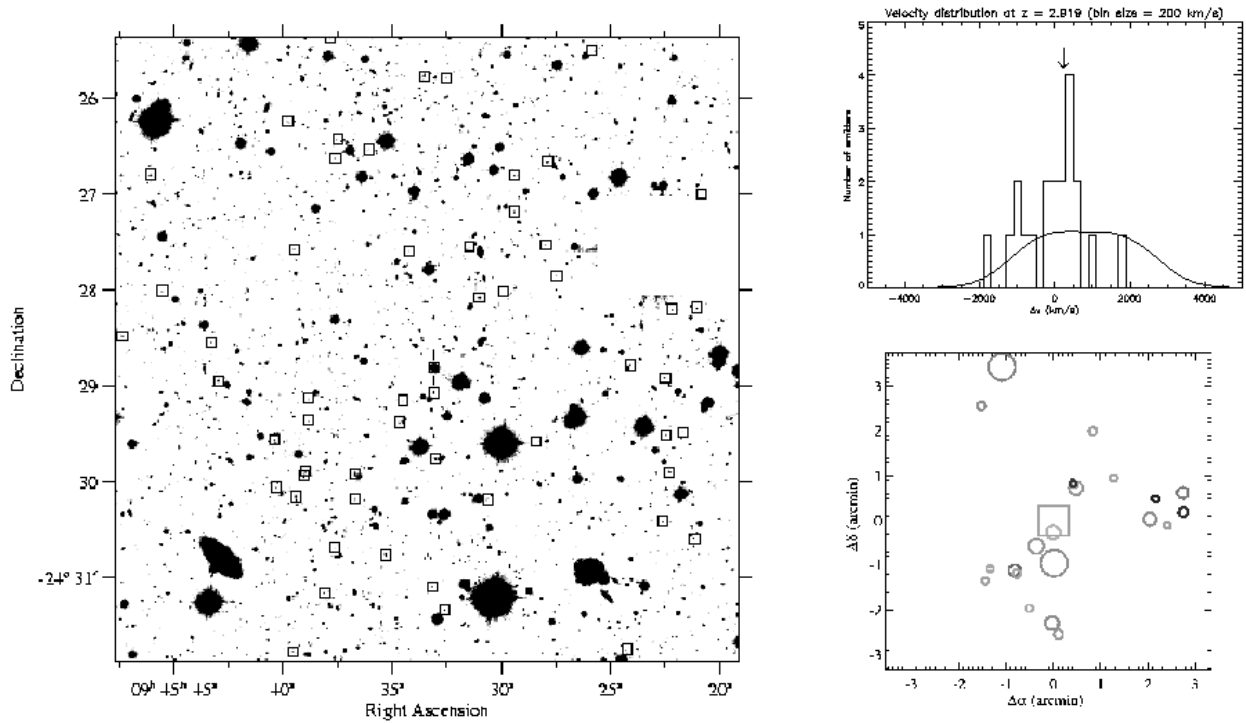}
\caption{
[Top Left] VLT narrow--band \lya\ image\cite{Venemans03} (grey scale; continuum not
subtracted) of the $z = 2.919$ radio galaxy MRC~0943$-$242 with LaEG's
identified (open squares). The blank area near the NW corner was occulted
by a slit because of a bright star.
[Top Right] Histogram of the velocity distribution of the 18 LaEG's +
radio galaxy which were spectroscopically confirmed.
[Bottom Right] Spatial distribution of the LaEG's (circles) + radio galaxy
(square).  Circle diameters are proportional to \lya\ flux  and range from
0.8 and 17 $\times 10^{-17}$ erg s$^{-1}$ cm$^{-2}$.
\label{fig:proto0943}}
\end{center}
\end{figure}

\mrc0943\ was imaged using a narrow--band filter with a 68~\AA\ bandpass
centered at 4781~\AA\ with the ESO 8.2m VLT Kueyen telescope using the
imaging mode of the FOcal Reducer/low dispersion Spectrograph 2 (FORS2).
The total exposure time was 22,500\,s, reaching a magnitude
limit of NB$_{AB}$ = 26.7 per  (3$\sigma$, 2.0\arcsec\ aperture) and
the data were obtained during photometric conditions and 0.8\arcsec\
seeing. A broad--band $B$ image (B$_{AB}$ = 27.0, 3$\sigma$, 2.0\arcsec\
aperture) was used to identify \lya\ excess galaxies in a manner as
described in previous papers\cite{Pentericci00,Venemans02}.
A total of 77 \lya\ excess galaxy candidates
were found above a limiting flux density of $5 \times 10^{-18} \rm
\,erg\,s^{-1}\,cm^{-2}\,arcsec^{-2}$ (3~$\sigma$; Fig 3).

We subsequently obtained multi--slit spectra of 24 of the LaEG candidates
with the upgraded dual beam Low Resolution Imaging Spectrograph\cite{Oke95}
at the Keck I 10m telescope. Two separate multi--slit masks were
used, with total exposure times of 10,800\,s for each.  We confirmed
that 18 of the 24 galaxies were at the redshift of the radio galaxy.
A histogram of the relative velocities and their spatial distribution is
shown in Figure 3. The velocity dispersion is large ($\sim$ 824 \kms)
with a hint of kinematic substructure at $\sim 1000$~\kms\ blueward of
the radio galaxy. A bimodal velocity distribution has also been found
in the $z = 2.156$ \hzrg\ proto--cluster around MRC~1138$-$262 (Table
2), and is fairly common even in much lower redshift galaxy clusters.
The spatial distribution shows possible evidence for substructure as
well, with small groups of 4 - 6 LaEGs to the West and South of the radio
galaxy. More spectroscopic observations of the remaining LaEG candidates
are planned to investigate this.

\subsection{\tn1338\ at $z = 4.102$}

\begin{figure}[t]
\begin{center}
\includegraphics[height=10cm]{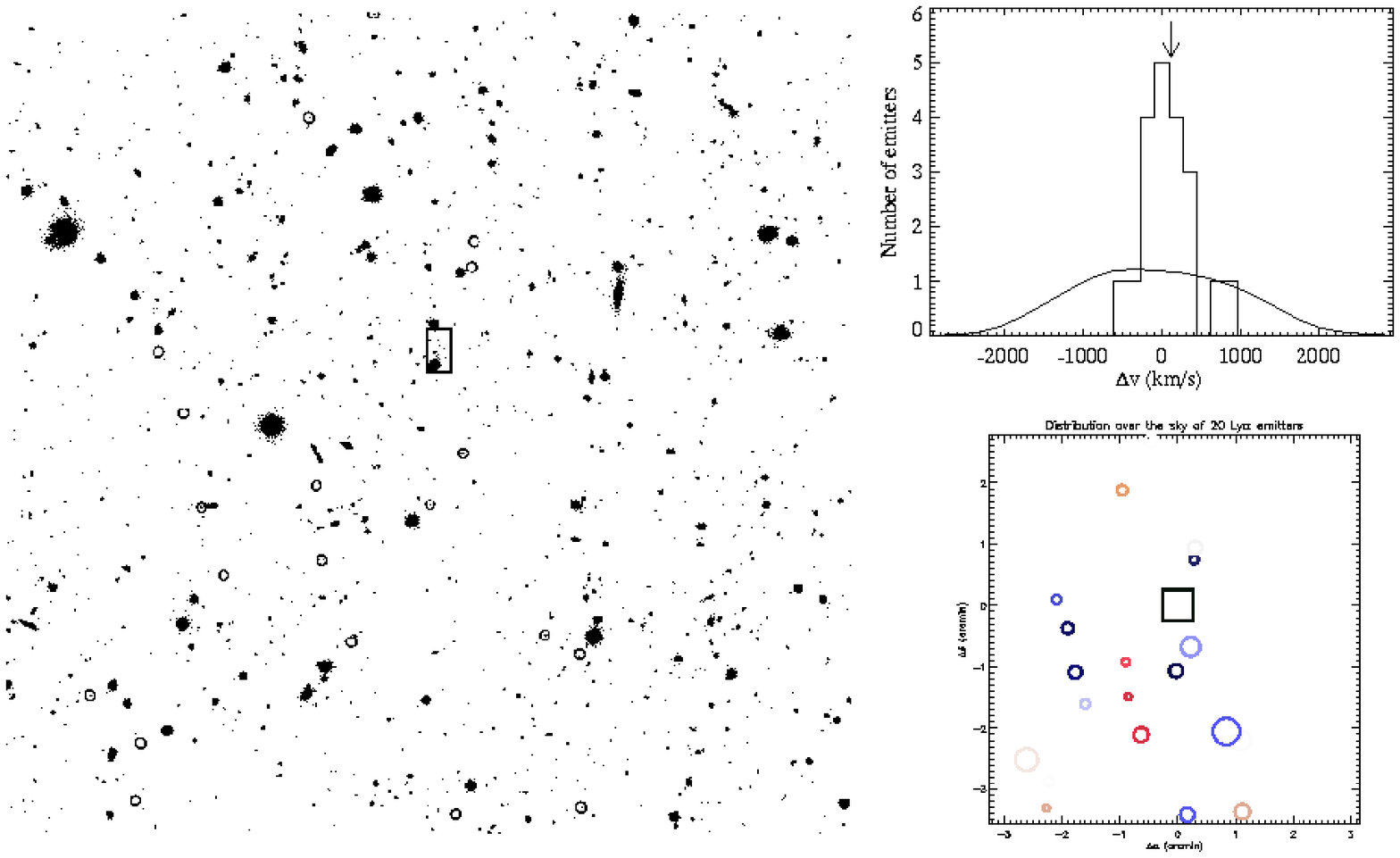}
\caption{
[Top Left] VLT narrow--band \lya\ image\cite{Venemans02} (grey scale; $6.4^\prime \times
6.2^\prime$) of \tn1338\ with 20 
spectroscopically confirmed LaEG candidates identified (circles).
[Top Right] 
Histogram of the velocity distribution of the LaEG's.
[Bottom Right] Spatial distribution of the LaEG's (circles) + radio galaxy
(square).  Circle diameters are proportional to \lya\ flux  and range from
0.8 and 17 $\times 10^{-17}$ erg s$^{-1}$ cm$^{-2}$.
\label{fig:proto1338}}
\end{center}
\end{figure}

\tn1338\ was imaged using a narrow--band filter with a 60~\AA\ bandpass
centered at the $z = 4.102$ redshifted \lya\ line (6202~\AA) with the
ESO VLT using the same telescope and instrument set up as for \mrc0943.
The total exposure time was 33,300\,s, reaching a magnitude limit
of NB$_{AB}$ = 27.4 (3$\sigma$, 2.0\arcsec\ aperture) and the data
were obtained during photometric conditions and 0.8\arcsec seeing. A
broad--band $B$ image (B$_{AB}$ = 27.02, 3$\sigma$, 2\arcsec\ aperture)
was used to identify LaEGS's in the same manner as for \mrc0943 .
A total of 33 objects were identified as potential LaEG's
(excluding the radio galaxy itself), 28 of which were retained after
rejecting 5 objects based on blue $B - R$ colors (see Ref.~\citenum{Venemans02} 
for details).

Multi--slit spectra of 23 LaEG candidates were obtained with the ESO
FORS2 spectrograph using two masks and $\sim$ 33,300\,s exposures each.
The observations, and Occam's razor\cite{Stern00}, confirmed
that 20 of the galaxies are at the redshift of \tn1338 (Fig. 4). The
velocity dispersion of the galaxies (326 \kms) is narrower than in the
other, lower redshift proto--clusters (Table 2) and shows no evidence for
substructure. The spatial distribution appears to be fairly random except
that the radio galaxy itself, which is presumably the most massive galaxy,
does {\it not} appear to be near the centroid of the proto--cluster.

\subsection{Conclusions for proto--clusters}

Narrow--band imaging of the fields around \hzrgs\ provides a powerful
method for finding proto--clusters over a large range of redshifts.
LaEG's represent only a small fraction ($20 - 25$\%) of the
total galaxy population in proto--clusters\cite{Steidel00}. However, for faint galaxies
($I_{AB} > 25.5$) at high redshifts emission--line objects such as these
may be the only ones for which redshifts may be measurable, even with 8 --
10m class telescopes.

The mass function of galaxy clusters and groups depends strongly on
cosmological parameters and evolves with lookback time\cite{Jenkins01}.
Observations of \hzrg\ proto--clusters will be very helpful in
investigating cosmological models in that they may be used to determine
the masses of galaxy clusters and groups at redshifts which can not be
reached by other methods (X--ray selection, gravitational lensing), or
which are prohibitive in observing time (`blind' multi--color surveys
with no prior redshift information).

The over--densities in the \hzrg\ proto--clusters are similar
to that in the SSA22$a$ over--dense region and lead to similar
estimates of the cluster masses of $10^{14} - 10^{15}$ \Msun.  In fact,
the luminosity functions and lifetimes of luminous radio sources are
consistent with {\it every} such over--dense field hosting a
massive galaxy that will be an active radio source at least 
once in its lifetime\cite{Venemans02}.

\section{Future work}
\label{sect:concl}

Large 8 -- 10m telescopes for the first time unlock the potential of
narrow--band imaging as a {\it cosmological} tool. Giant emission--line
nebulae are now known to exist out to $z \sim 4$, some with obvious
ionizing sources, some without (Table 1). By studying their properties we
may learn about the formation of galaxies, the feedback from AGN/starburst
winds, and the chemical enrichment of intra--cluster media.  Radio--loud
galaxies seem endowed by rather spectacular nebulae, probably they are
the most massive forming galaxies, but other types of objects need to
be studied as well. For example, if quasars would become active during
the very early stages of galaxy formation, then one might expect them
to boost the \lya\ emission in the surrounding gas, which might be used
to learn more about the galaxy formation process\cite{Haiman01}.

Emission--line imaging of the fields around targeted high redshift
objects can push searches for galaxy proto--clusters to much higher
redshifts than would otherwise be possible. Again radio galaxies are
ideally suited for this because they live in the most massive galaxies
which live in the most over--dense regions.

To fully exploit emission--line imaging as a cosmological tool
will require large field--of--view ($0.5 \times 0.5$ sq.deg) 
tunable filters\cite{Bland01}, Imaging Fourier Transform
Spectrometers\cite{Wurtz02}, or other types of multi--wavelength `3--D' imaging
devices\cite{vanBreugel00}.  Large ($>~200$~cm$^2$)
custom made narrow--band filters are expensive, and it is an unfortunate
fact that the largest telescopes also require the largest filters,
even for modest sized fields of view (273~cm$^2$ for a 81~sq.arcmin 
FOV in the case
of DEIMOS at Keck for example). This problem will be even more acute
for the next generation of `extremely' large telescopes such as the
CELT~30m. It is also clear that the selection of emission--line objects
provides only partial insight into the galaxy populations and kinematics
of proto--clusters, and obtaining spectroscopic absorption line redshifts
of faint galaxies will be very difficult.  The most powerful observational
tool here would again be a 3--D imaging device with selectable, medium
sized (few 100\AA\ wide) bandpasses so that accurate photometric redshift
measurements can be made.


\acknowledgments     
 
The work by WvB, MR, WdV, and AS was performed under the auspices of
the US Department of Energy under contract W-7405-ENG-48.  W.v.B.\ also
acknowledges NASA grants GO~5940, 6608 and 8183 in support
of \hzrg\ research with HST. WvB is greatful for enlightening discussions
with D. Mathiesen (LLNL) on the use of proto--cluster observations for
constraining cosmological parameters, and to J.~Reed for inspiration.


\bibliography{spie}   

\begin{thebibliography}{10}

\bibitem{Rees78}
M.~J. {Rees}, ``{Relativistic jets and beams in radio galaxies},'' {\em \nat}
  {\bf 275}, pp.~516--+, Oct. 1978.

\bibitem{Blandford82}
R.~D. {Blandford} and D.~G. {Payne}, ``{Hydromagnetic flows from accretion
  discs and the production of radio jets},'' {\em \mnras} {\bf 199},
  pp.~883--903, June 1982.

\bibitem{Magorrian98}
J.~{Magorrian}, S.~{Tremaine}, D.~{Richstone}, R.~{Bender}, G.~{Bower},
  A.~{Dressler}, S.~M. {Faber}, K.~{Gebhardt}, R.~{Green}, C.~{Grillmair},
  J.~{Kormendy}, and T.~{Lauer}, ``{The Demography of Massive Dark Objects in
  Galaxy Centers},'' {\em \aj} {\bf 115}, pp.~2285--2305, June 1998.

\bibitem{Gebhardt00}
K.~{Gebhardt}, R.~{Bender}, G.~{Bower}, A.~{Dressler}, S.~M. {Faber}, A.~V.
  {Filippenko}, R.~{Green}, C.~{Grillmair}, L.~C. {Ho}, J.~{Kormendy}, T.~R.
  {Lauer}, J.~{Magorrian}, J.~{Pinkney}, D.~{Richstone}, and S.~{Tremaine},
  ``{A Relationship between Nuclear Black Hole Mass and Galaxy Velocity
  Dispersion},'' {\em \apjl} {\bf 539}, pp.~L13--LL16, Aug. 2000.

\bibitem{Ferrarese00}
L.~{Ferrarese} and D.~{Merritt}, ``{A Fundamental Relation between Supermassive
  Black Holes and Their Host Galaxies},'' {\em \apjl} {\bf 539}, pp.~L9--LL12,
  Aug. 2000.

\bibitem{DeBreuck02a}
C.~{De Breuck}, W.~{van Breugel}, S.~A. {Stanford}, H.~{R{\" o}ttgering},
  G.~{Miley}, and D.~{Stern}, ``{Optical and Near-Infrared Imaging of
  Ultra-Steep-Spectrum Radio Sources: The K-z Diagram of Radio-selected and
  Optically Selected Galaxies},'' {\em \aj} {\bf 123}, pp.~637--677, Feb. 2002.

\bibitem{Jarvis01}
M.~J. {Jarvis}, S.~{Rawlings}, S.~{Eales}, K.~M. {Blundell}, A.~J. {Bunker},
  S.~{Croft}, R.~J. {McLure}, and C.~J. {Willott}, ``{A sample of 6C radio
  sources designed to find objects at redshift z $>$ 4 - III. Imaging and the
  radio galaxy K-z relation},'' {\em \mnras} {\bf 326}, pp.~1585--1600, Oct.
  2001.

\bibitem{Dey97}
A.~{Dey}, W.~{van Breugel}, W.~D. {Vacca}, and R.~{Antonucci}, ``{Triggered
  Star Formation in a Massive Galaxy at Z = 3.8: 4C 41.17},'' {\em \apj} {\bf
  490}, pp.~698--+, Dec. 1997.

\bibitem{Pentericci99}
L.~{Pentericci}, H.~J.~A. {R{\" o}ttgering}, G.~K. {Miley}, P.~{McCarthy},
  H.~{Spinrad}, W.~J.~M. {van Breugel}, and F.~{Macchetto}, ``{HST images and
  properties of the most distant radio galaxies},'' {\em \aap} {\bf 341},
  pp.~329--347, Jan. 1999.

\bibitem{Archibald01}
E.~N. {Archibald}, J.~S. {Dunlop}, D.~H. {Hughes}, S.~{Rawlings}, S.~A.
  {Eales}, and R.~J. {Ivison}, ``{A submillimetre survey of the star formation
  history of radio galaxies},'' {\em \mnras} {\bf 323}, pp.~417--444, May 2001.

\bibitem{Reuland02}
M.~{Reuland}, W.~{van Breugel}, H.~{R\"ottgering}, d.~W., S.~{Stanford},
  A.~{Dey}, M.~{Lacy}, J.~{Bland-Hawthorn}, M.~{Dopita}, and M.~{Miley},
  ``{Giant Ly-a nebulae associated with massive forming galaxies},'' {\em \apj}
  {\bf 000}, p.~000, submitted 2002.

\bibitem{Papadopoulos00}
P.~P. {Papadopoulos}, H.~J.~A. {R{\" o}ttgering}, P.~P. {van der Werf},
  S.~{Guilloteau}, A.~{Omont}, W.~J.~M. {van Breugel}, and R.~P.~J. {Tilanus},
  ``{CO (4-3) and Dust Emission in Two Powerful High-Z Radio Galaxies, and CO
  Lines at High Redshifts},'' {\em \apj} {\bf 528}, pp.~626--636, Jan. 2000.

\bibitem{DeBreuck02b}
C.~{De Breuck}, R.~{Neri}, A.~{Omont}, R.~{Morganti}, B.~{Rocca-Volmerage},
  M.~{Reuland}, D.~{Stern}, D.~{Stevens}, W.~{van Breugel}, H.~{R{\"
  o}ttgering}, S.~{Stanford}, M.~{Vigotti}, H.~{Spinrad}, and M.~{Wright},
  ``{CO emission and associated HI absorption from a massive gas reservoir
  surrounding the z = 3 radio galaxy B3~J2330+3927},'' {\em \aap} {\bf 000},
  p.~000, submitted 2002.

\bibitem{Kaiser84}
N.~{Kaiser}, ``{On the spatial correlations of Abell clusters},'' {\em \apjl}
  {\bf 284}, pp.~L9--LL12, Sept. 1984.

\bibitem{Eisenhardt92}
P.~{Eisenhardt} and M.~{Dickinson}, ``{How old is the Z = 3.4 radio galaxy B2
  0902+34?},'' {\em \apjl} {\bf 399}, pp.~L47--LL50, Nov. 1992.

\bibitem{Chambers90}
K.~C. {Chambers}, G.~K. {Miley}, and W.~J.~M. {van Breugel}, ``{4C 41.17 - A
  radio galaxy at a redshift of 3.8},'' {\em \apj} {\bf 363}, pp.~21--39, Nov.
  1990.

\bibitem{vanOjik96}
R.~{van Ojik}, H.~J.~A. {Roettgering}, C.~L. {Carilli}, G.~K. {Miley}, M.~N.
  {Bremer}, and F.~{Macchetto}, ``{A powerful radio galaxy at z=3.6 in a giant
  rotating Ly-a halo.},'' {\em \aap} {\bf 313}, pp.~25--44, Sept. 1996.

\bibitem{Kurk02}
J.~D. {Kurk}, L.~{Pentericci}, H.~J.~A. {R{\" o}ttgering}, and G.~K. {Miley},
  ``{Observations of Radio Galaxy MRC 1138-262: Merging Galaxies Embedded in a
  Giant Ly-a Halo},'' in {\em Revista Mexicana de Astronomia y Astrofisica
  Conference Series},   {\bf 13}, pp.~191--195, May 2002.

\bibitem{Francis01}
P.~J. {Francis}, G.~M. {Williger}, N.~R. {Collins}, P.~{Palunas}, E.~M.
  {Malumuth}, B.~E. {Woodgate}, H.~I. {Teplitz}, A.~{Smette}, R.~S.
  {Sutherland}, A.~C. {Danks}, R.~S. {Hill}, D.~{Lindler}, R.~A. {Kimble},
  S.~R. {Heap}, and J.~B. {Hutchings}, ``{A Pair of Compact Red Galaxies at
  Redshift 2.38, Immersed in a 100 Kiloparsec Scale Ly-a Nebula},'' {\em \apj}
  {\bf 554}, pp.~1001--1011, June 2001.

\bibitem{Steidel00}
C.~C. {Steidel}, K.~L. {Adelberger}, A.~E. {Shapley}, M.~{Pettini},
  M.~{Dickinson}, and M.~{Giavalisco}, ``{Ly-a Imaging of a Proto-Cluster
  Region at z=3.09},'' {\em \apj} {\bf 532}, pp.~170--182, Mar. 2000.

\bibitem{Venemans02}
B.~P. {Venemans}, J.~D. {Kurk}, G.~K. {Miley}, H.~J.~A. {R{\" o}ttgering},
  W.~{van Breugel}, C.~L. {Carilli}, C.~{De Breuck}, H.~{Ford}, T.~{Heckman},
  P.~{McCarthy}, and L.~{Pentericci}, ``{The Most Distant Structure of Galaxies
  Known: A Protocluster at z=4.1},'' {\em \apjl} {\bf 569}, pp.~L11--LL14, Apr.
  2002.

\bibitem{Sheinis00}
A.~I. {Sheinis}, J.~S. {Miller}, M.~{Bolte}, and B.~M. {Sutin}, ``{Performance
  characteristics of the new Keck Observatory echelle spectrograph and
  imager},'' in {\em Proc. SPIE Vol. 4008, p. 522-533, Optical and IR Telescope
  Instrumentation and Detectors, Masanori Iye; Alan F. Moorwood; Eds.},   {\bf
  4008}, pp.~522--533, Aug. 2000.

\bibitem{vanBreugel86}
W.~J.~M. {van Breugel}, T.~M. {Heckman}, G.~K. {Miley}, and A.~V. {Filippenko},
  ``{4C 29.30 - Extended optical line and radio emission in a probable galaxy
  merger},'' {\em \apj} {\bf 311}, pp.~58--84, Dec. 1986.

\bibitem{Dopita02}
M.~A. {Dopita}, B.~A. {Groves}, R.~S. {Sutherland}, L.~{Binette}, and
  G.~{Cecil}, ``{Are the Narrow-Line Regions in Active Galaxies Dusty and
  Radiation Pressure Dominated?},'' {\em \apj} {\bf 572}, pp.~753--761, June
  2002.

\bibitem{Lehnert99}
M.~D. {Lehnert}, T.~M. {Heckman}, and K.~A. {Weaver}, ``{Very Extended X-Ray
  and H-a Emission in M82: Implications for the Superwind Phenomenon},'' {\em
  \apj} {\bf 523}, pp.~575--584, Oct. 1999.

\bibitem{McLean98}
I.~S. {McLean}, E.~E. {Becklin}, O.~{Bendiksen}, G.~{Brims}, J.~{Canfield},
  D.~F. {Figer}, J.~R. {Graham}, J.~{Hare}, F.~{Lacayanga}, J.~E. {Larkin},
  S.~B. {Larson}, N.~{Levenson}, N.~{Magnone}, H.~{Teplitz}, and W.~{Wong},
  ``{Design and development of NIRSPEC: a near-infrared echelle spectrograph
  for the Keck II telescope},'' in {\em Proc. SPIE Vol. 3354, p. 566-578,
  Infrared Astronomical Instrumentation, Albert M. Fowler; Ed.},   {\bf 3354},
  pp.~566--578, Aug. 1998.

\bibitem{Carilli88}
C.~L. {Carilli}, R.~A. {Perley}, and J.~H. {Dreher}, ``{Discovery of the bow
  shock of Cygnus A},'' {\em \apjl} {\bf 334}, pp.~L73--LL76, Nov. 1988.

\bibitem{Begelman89}
M.~C. {Begelman} and D.~F. {Cioffi}, ``{Overpressured cocoons in extragalactic
  radio sources},'' {\em \apjl} {\bf 345}, pp.~L21--LL24, Oct. 1989.

\bibitem{Maxfield02}
L.~{Maxfield}, H.~{Spinrad}, D.~{Stern}, A.~{Dey}, and M.~{Dickinson},
  ``{Spatially Resolved High-Ionization Nebulae in a High-Redshift Radio
  Galaxy: Evidence for Shock Ionization and Photoionization},'' {\em \aj} {\bf
  123}, pp.~2321--2332, May 2002.

\bibitem{Binette00}
L.~{Binette}, J.~D. {Kurk}, M.~{Villar-Mart{\' i}n}, and H.~J.~A. {R{\"
  o}ttgering}, ``{A vestige low metallicity gas shell surrounding the radio
  galaxy 0943-242 at z=2.92},'' {\em \aap} {\bf 356}, pp.~23--32, Apr. 2000.

\bibitem{Jarvis02}
M.~{Jarvis}, R.~{Wilman}, H.~{R\"ottgering}, and L.~{Binette}, ``{Probing the
  absorbing haloes around two high-redshift radio galaxies with VLT-UVES},''
  {\em \mnras} {\bf 000}, p.~000, submitted 2002.

\bibitem{McCarthy87}
P.~J. {McCarthy}, W.~{van Breugel}, H.~{Spinrad}, and S.~{Djorgovski}, ``{A
  correlation between the radio and optical morphologies of distant 3Cr radio
  galaxies},'' {\em \apjl} {\bf 321}, pp.~L29--LL33, Oct. 1987.

\bibitem{Bicknell00}
G.~V. {Bicknell}, R.~S. {Sutherland}, W.~J.~M. {van Breugel}, M.~A. {Dopita},
  A.~{Dey}, and G.~K. {Miley}, ``{Jet-induced Emission-Line Nebulosity and Star
  Formation in the High-Redshift Radio Galaxy 4C 41.17},'' {\em \apj} {\bf
  540}, pp.~678--686, Sept. 2000.

\bibitem{Chapman01}
S.~C. {Chapman}, G.~F. {Lewis}, D.~{Scott}, E.~{Richards}, C.~{Borys}, C.~C.
  {Steidel}, K.~L. {Adelberger}, and A.~E. {Shapley}, ``{Submillimeter Imaging
  of a Protocluster Region at Z=3.09},'' {\em \apjl} {\bf 548}, pp.~L17--LL21,
  Feb. 2001.

\bibitem{Fardal01}
M.~A. {Fardal}, N.~{Katz}, J.~P. {Gardner}, L.~{Hernquist}, D.~H. {Weinberg},
  and R.~{Dav{\' e}}, ``{Cooling Radiation and the Ly-a Luminosity of Forming
  Galaxies},'' {\em \apj} {\bf 562}, pp.~605--617, Dec. 2001.

\bibitem{Haiman00}
Z.~{Haiman}, M.~{Spaans}, and E.~{Quataert}, ``{Ly-a Cooling Radiation from
  High-Redshift Halos},'' {\em \apjl} {\bf 537}, pp.~L5--LL8, July 2000.

\bibitem{Silk87}
J.~{Silk}, ``{Galaxy formation},'' in {\em IAU Symp. 117: Dark matter in the
  universe},   {\bf 117}, pp.~335--354, 1987.

\bibitem{Silk98}
J.~{Silk} and M.~J. {Rees}, ``{Quasars and galaxy formation},'' {\em \aap} {\bf
  331}, pp.~L1--LL4, Mar. 1998.

\bibitem{Nulsen02}
P.~E.~J. {Nulsen}, L.~P. {David}, B.~R. {McNamara}, C.~{Jones}, W.~R. {Forman},
  and M.~{Wise}, ``{Interaction of Radio Lobes with the Hot Intracluster
  Medium: Driving Convective Outflow in Hydra A},'' {\em \apj} {\bf 568},
  pp.~163--173, Mar. 2002.

\bibitem{Kaiser01}
M.~{Br{\" u}ggen} and C.~R. {Kaiser}, ``{Buoyant radio plasma in clusters of
  galaxies},'' {\em \mnras} {\bf 325}, pp.~676--684, Aug. 2001.

\bibitem{Odea02}
C.~P. {O'Dea}, ``{Life cycles of radio galaxies: introductory remarks},'' {\em
  New Astronomy Review} {\bf 46}, pp.~41--46, May 2002.

\bibitem{Pentericci00}
L.~{Pentericci}, J.~D. {Kurk}, H.~J.~A. {R{\" o}ttgering}, G.~K. {Miley},
  W.~{van Breugel}, C.~L. {Carilli}, H.~{Ford}, T.~{Heckman}, P.~{McCarthy},
  and A.~{Moorwood}, ``{A search for clusters at high redshift. II. A proto
  cluster around a radio galaxy at z=2.16},'' {\em \aap} {\bf 361},
  pp.~L25--LL28, Sept. 2000.

\bibitem{Keel99}
W.~C. {Keel}, S.~H. {Cohen}, R.~A. {Windhorst}, and I.~{Waddington},
  ``{Evidence for Large-Scale Structure at z \~{} 2.4 from Ly-a Imaging},''
  {\em \aj} {\bf 118}, pp.~2547--2560, Dec. 1999.

\bibitem{Venemans03}
B.~P. {Venemans} and {et. al.}, ``{A Protocluster around the z = 2.919 radio
  galaxy MRC~0943$-$242},'' {\em \aap} {\bf 000}, p.~000, in preparation 2003.

\bibitem{Stanford97}
S.~A. {Stanford}, R.~{Elston}, P.~R. {Eisenhardt}, H.~{Spinrad}, D.~{Stern},
  and A.~{Dey}, ``{An IR-Selected Galaxy Cluster at z=1.27},'' {\em \aj} {\bf
  114}, pp.~2232--+, Dec. 1997.

\bibitem{vanDokkum01}
P.~G. {van Dokkum}, S.~A. {Stanford}, B.~P. {Holden}, P.~R. {Eisenhardt},
  M.~{Dickinson}, and R.~{Elston}, ``{The Galaxy Population of Cluster RX
  J0848+4453 at Z=1.27},'' {\em \apjl} {\bf 552}, pp.~L101--LL104, May 2001.

\bibitem{Steidel98}
C.~C. {Steidel}, K.~L. {Adelberger}, M.~{Dickinson}, M.~{Giavalisco},
  M.~{Pettini}, and M.~{Kellogg}, ``{A Large Structure of Galaxies at Redshift
  Z approximately 3 and Its Cosmological Implications},'' {\em \apj} {\bf 492},
  pp.~428--+, Jan. 1998.

\bibitem{Oke95}
J.~B. {Oke}, J.~G. {Cohen}, M.~{Carr}, J.~{Cromer}, A.~{Dingizian}, F.~H.
  {Harris}, S.~{Labrecque}, R.~{Lucinio}, W.~{Schaal}, H.~{Epps}, and
  J.~{Miller}, ``{The Keck Low-Resolution Imaging Spectrometer},'' {\em \pasp}
  {\bf 107}, pp.~375--+, Apr. 1995.

\bibitem{Stern00}
D.~{Stern}, A.~{Bunker}, H.~{Spinrad}, and A.~{Dey}, ``{One-Line Redshifts and
  Searches for High-Redshift Ly-a Emission},'' {\em \apj} {\bf 537},
  pp.~73--79, July 2000.

\bibitem{Jenkins01}
A.~{Jenkins}, C.~S. {Frenk}, S.~D.~M. {White}, J.~M. {Colberg}, S.~{Cole},
  A.~E. {Evrard}, H.~M.~P. {Couchman}, and N.~{Yoshida}, ``{The mass function
  of dark matter haloes},'' {\em \mnras} {\bf 321}, pp.~372--384, Feb. 2001.

\bibitem{Haiman01}
Z.~{Haiman} and M.~J. {Rees}, ``{Extended Ly-a Emission around Young Quasars: A
  Constraint on Galaxy Formation},'' {\em \apj} {\bf 556}, pp.~87--92, July
  2001.

\bibitem{Bland01}
J.~{Bland-Hawthorn}, W.~{van Breugel}, P.~R. {Gillingham}, I.~K. {Baldry}, and
  D.~H. {Jones}, ``{A Tunable Lyot Filter at Prime Focus: a Method for Tracing
  Supercluster Scales at z \~{} 1},'' {\em \apj} {\bf 563}, pp.~611--628, Dec.
  2001.

\bibitem{Wurtz02}
R.~{Wurtz}, E.~{Wishnow}, S.~{Blais-Ouellette}, K.~{Cook}, D.~{Carr},
  F.~{Grandmont}, I.~{Lewis}, and C.~{Stubbs}, ``{LIFTS: an Imaging Fourier
  Transform Spectrograph for Astronomy},'' in {\em Presented at "Galaxies: the
  Third Dimension" held in Cozumel, Mexico, in December 2001, eds. M. Rosado,
  L. Binette, L. Arias, ASP Conf. Ser},  pp.~3501--+, Mar. 2002.

\bibitem{vanBreugel00}
W.~{van Breugel} and J.~{Bland-Hawthorn}, eds., {\em {Imaging the Universe in
  Three Dimensions, ASP Conf. Ser. 195}}, 2000.

\end{thebibliography}
\bibliographystyle{spiebib}   


\end{document}